\documentclass[%
 reprint,
 amsmath,amssymb,
 aps,
]{revtex4-2}

\usepackage{graphicx}% Include figure files
\usepackage{dcolumn}% Align table columns on decimal point
\usepackage{bm}% bold math
\usepackage{physics}
\usepackage{slashed}
\usepackage{mathtools}
\begin{document}

\preprint{APS/123-QED}

\title{Novel Constraints on Spin-Dependent Light Dark Matter Scattering}% Force line breaks with \\

\author{$^{1}$Alex Clarke}%Lines break automatically or can be forced with \\
\author{$^{1,2}$Maxim Pospelov}%
\affiliation{%
$^{1}$School of Physics and Astronomy, University of Minnesota, Minneapolis, MN 55455, USA 
}%
\affiliation{%
$^{2}$William I. Fine Theoretical Physics Institute, School of Physics and Astronomy,
University of Minnesota, Minneapolis, MN 55455, USA
}%

\date{\today}% It is always \today, today,

\begin{abstract}
We explore the sensitivity of the SNO experiment to light dark matter particles $\chi$ with spin-dependent interactions with nucleons. We show that the pair-production of MeV scale dark matter is possible in heavy water (CANDU) reactors via ${\rm D}(n,\chi\bar\chi)^3{\rm He}$, and calculate the expected rate within the simplest models of $\chi$-nucleon interactions. 
%Heavy water nuclear reactors serve as an excellent production method for spin-dependent dark matter. 
Owing to a sizable $Q$-value for this reaction, a large fraction of DM particles produced this way are above the threshold for deuteron disintegration, ${\rm D}(\chi,\chi)np$, which adds to the SNO neutral current signal. Evaluating the CANDU-to-SNO scheme for the production and detection of DM, we derive novel constraints for the $\chi$-nucleon spin-dependent cross sections, showing that cross sections above $\sigma_{\chi p} \sim 10^{-33}\,{\rm cm}^{2}$ are generally excluded if $m_\chi \leq1.5$\,MeV. An isospin-mirror reaction will occur in the Sun, and for the kinematically allowed region it excludes a portion of parameter space with cross sections on the order $10^{-37}\,{\rm cm}^{2}$. We also evaluate the potential sensitivity of small ``near" detectors placed in close proximity to a CANDU reactor to search for a coherent nuclear recoil, finding subdominant sensitivity. 
\end{abstract}

\maketitle

%\tableofcontents

\section{\label{sec:level1}Introduction}
A wealth of cosmological and astrophysical observations point towards the existence of a non-baryonic component to the energy density in the universe known as dark matter (DM) \cite{BERTONE2005279}. Despite immense experimental and theoretical effort, the exact nature of DM is still unknown. DM candidates, called $\chi$ in this paper, occupy a large space of possibilities with particle DM mass, $m_\chi$, and DM interaction strengths with the Standard Model particles being the most experimentally relevant. For suitable choices of these parameters, physicists pursue the DM direct detection avenue, consisting of searches for small amounts of energy deposited by DM scattering with atoms in high sensitivity detectors. While no reproducible positive signals from $\chi$ scattering on atoms were found, the program of direct detection is nevertheless a story of dramatic progress that produced stringent constraints on DM scattering cross sections with nucleons and electrons, $\sigma_{\chi n}$ and $\sigma_{\chi e}$, eliminating some of the most straightforward options in the process \cite{Lee:1977ua}. The direction in the parameter space towards the smallest possible $\sigma_{\chi n}$ is limited by the amount of target material in these detectors, and by the background levels.  For $m_\chi$ below 1 GeV, the sensitivity to $\sigma_{\chi n}$ worsens dramatically, because of tiny recoil energy and exceedingly large backgrounds. Alternate detection methods are required to circumvent this limitation in sensitivity.   

 Several strategies can be chosen to improve sensitivity to light DM scattering on nucleons. For example, new sensitive devices can be built where the energy thresholds can be $O({\rm eV})$ (see, {\em e.g.} recent paper \cite{TESSERACT:2025tfw}). If backgrounds can be kept under control, it is conceivable that new detection methods could bring $m_\chi$ thresholds down below tens of MeV. 

One can circumvent the limitation of small $m_\chi$ by considering the inelastic collisions of DM with nuclei. If there exists a DM fraction that has sufficient energy to excite/break-up a nucleus, then the resulting products can produce a signal which exceeds sensitivity for elastic collisions with nuclei. Exploiting this increase in sensitivity, the parameters of DM in this mass range can be further constrained. One way of creating a ``velocitized" fraction of total DM flux is by exploring multiple collisions \cite{Bringmann:2018cvk}. For example, first collisions with galactic cosmic rays (CR) can create a fast fraction of DM that then can scatter on a nucleus with a sufficient amount of energy to spare. This results in constraints on the DM cross section over a wide range of $m_\chi$, and for both spin-dependent (SD) and spin-independent (SI) $\sigma_{\chi n}$ \cite{Bringmann:2018cvk}. Another possibility is to accept strong inefficiency in passing the nuclear recoil to electrons via the so-called Migdal effect, gaining a much more robust energy release in the process of atomic ionization (see {\em e.g.} Refs. \cite{Kouvaris:2016afs,Ibe:2017yqa}). The focus of this paper is the inelastic interaction of energetic DM particles with deuterons, which is experimentally relevant, and theoretically fairly straightforward. DM-induced nuclear inelastic processes have been considered in a variety of settings before, see Refs.\,\cite{Engel:1999kv,Pospelov:2011ha,Baudis:2013bba,Dutta:2023fij}. 

Light DM can be indeed pair-produced in collisions or decays of metastable SM particles. One could then search for an abnormal amount of missing energy taken by the DM particles \cite{Bird:2004ts,Gninenko:2014pea,Izaguirre:2014bca}. If an intense source of light DM is created, one could hope for a second collision of DM inside a detector, which would then constitute the appearance of a signal \cite{Batell:2009di,Izaguirre:2013uxa,Batell:2014mga}. If the mass of exotic particles is in the MeV range, one should explore the most intense sources of MeV quanta, namely commercial nuclear reactors, as possible DM pair-production sites. Recently, it was shown that nuclear reactors can provide the most sensitive bounds on axion-like particles, milli-charged particles, and other exotics \cite{Brdar:2020dpr,NEON:2024kwv,CONNIE:2024off,Gao:2025ykc}. Additional impetus for considering reactors as sources of MeV DM is provided by the experimental programs that pursue the measurement of the elastic neutrino-nucleus scattering by placing sensitive detectors (developed for DM detection) in close proximity to reactor cores. Last year has seen the first successful measurement of such a process (sometimes referred to as CE$\nu$NS) with the use of a well-shielded low-threshold Ge detector in close proximity to a commercial nuclear reactor \cite{Ackermann:2025obx}. In addition to reactors, the Sun represents a very intense source of nuclear processes capable of pair-producing MeV-scale DM. In that case, the sensitivity would be regulated by the solar opacity, which would not allow DM particles with substantial scattering cross sections to leave the Sun without losing most of their energy (and therefore rendering them below the detection threshold). 

\begin{figure}[t]
\includegraphics[width=0.44\textwidth]{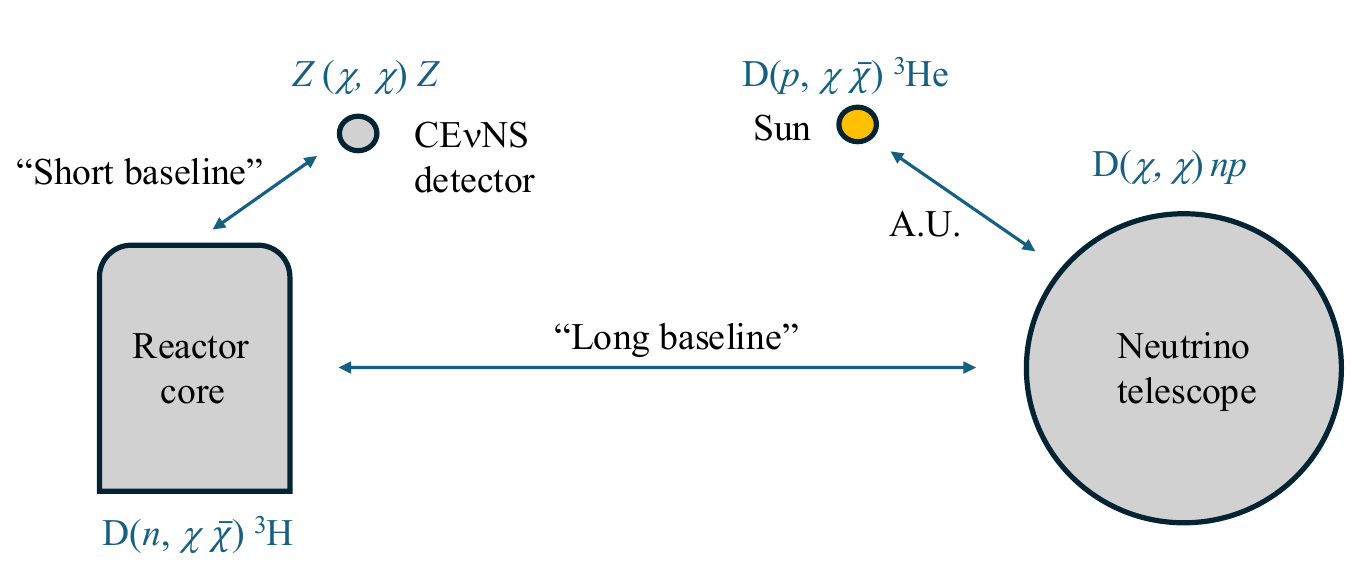}% Here is how to import EPS art
\caption{ \label{CANDUtoSNO} A not-to-scale depiction of emission/detection scheme of MeV-scale DM. $\chi\bar\chi$ pairs are emitted in a reactor core, and we concentrate on those produced in heavy water reactors. $\chi\bar\chi$ pairs are also emitted in the sun. Detectors purposely installed in proximity $O(30\,\rm m)$ to reactors (short baseline) may be used to detect $\chi\bar\chi$ pairs produced in reactors, while neutrino telescopes, such as SNO at long baseline $O(200\,\rm km)$ to reactors may be used to detect $\chi\bar\chi$ pairs produced in reactors and the sun. Representative reactions for the pair-creation of DM and subsequent detection are shown in blue.}
\end{figure}

In this paper, we consider the sensitivity to the light DM SD scattering on nucleons, starting from the effective spin-spin interaction. We evaluate the production of MeV scale DM in nuclear reactors, finding that heavy water, D$_2$O, reactors can generate an appreciable amount of $\chi\bar\chi$ pairs, with energy extending to several MeVs. In addition, we evaluate the production of $\chi\bar\chi$ pairs in the Sun in the main $pp$ chain of reactions. We then evaluate various detection mechanisms, that include SD scattering of newly produced DM particles in the near detector, and the scattering signatures in neutrino telescopes. The schematic idea for our study is illustrated in Fig.\,\ref{CANDUtoSNO}. 

Specifically, we evaluate the probability of SD pair-production of DM in nuclear reactors. Since we study SD interactions, we should concentrate on channels for $\chi \bar\chi$ emission analogous to $M1$ $\gamma$-emission. For natural water reactors, one of the dominant $M1$ channels is neutron capture on protons, $p(n,\gamma)\rm D$, which releases only 2.2\,MeV of energy, which turns out to be rather insufficient for most subsequent applications, as {\em e.g.} elastic scattering on nuclei require larger initial energies for detecting $\chi$ (or $\bar \chi$) above the detector threshold. Another major ``sink" of neutrons is the $(n,\gamma)$ reaction on $^{238}\rm U$, which should have some fraction of $M1$ transitions but is dominated by $E1$ emission. It is interesting to note that in D$_2$O the corresponding reaction of neutron capture is given by 
\begin{eqnarray}
\label{R1}
R1&:& ~~   \text{D} +n \to {\rm ^3H} + \gamma\\
R2&:& ~~   \text{D} +n \to {\rm ^3H} + (\chi\bar{\chi}),
\label{R2}
\end{eqnarray}
and $R1$ releases $Q(R1) \simeq 6.257 \,\rm MeV$ of energy. Our goal is to evaluate the ratio of the rates as functions of the input $\chi$-nucleon cross section, $R2/R1(\sigma_{\chi n})$. Since the occurrence of $R1$ is known, this would allow us to calculate the SD DM production intensity. Other important sources of spin-dependent emission include neutron capture on reactor walls, ${\rm ^{56} Fe}(n,\gamma){\rm ^{57} Fe}$, which would occur in both natural and heavy water reactors. The energy release in this reaction is $\sim 7.6\,\rm MeV$. Some of these reactions were already addressed in the context of neutron-capture-initiated axion emission in the planned beam dump experiments \cite{Waites:2022tov}. 

The {\em isospin-mirror} of $R1$ and $R2$ will occur in the Sun, as part of the main $pp$ cycle of energy generation,
\begin{eqnarray}
\label{R3}
R3&:& ~~   \text{D} +p \to {\rm ^3He} + \gamma\\
R4&:& ~~   \text{D} +p \to {\rm ^3He} + (\chi\bar{\chi})
\label{R4}
\end{eqnarray}
The $Q$-value of this reaction is slightly smaller, $Q(R3)= 5.49$\,MeV. Therefore, the Sun can also be a source of MeV-scale ``boosted" or ``velocitized" DM particles.  

Detection of $\chi$($\bar\chi$) particles emitted in the core of a reactor (or the solar core) could occur via different methods. Some large neutrino telescopes can be located at $O(100\,\rm km)$ distance from the reactors \cite{KamLAND:2002uet,SNO:2003bmh,JUNO:2025gmd}. Scattering of $\chi$'s inside large volumes of hydrocarbon or water-based detectors could result in nuclear excitations that may be visible above the background (See Refs. \cite{Dutta:2022tav,Bell:2022dbf} for alternative inelastic channels). For this paper, we concentrate on the most relevant SD process for the reactor and solar source that will occur in a D$_2$O-based detector:
\begin{eqnarray}
R5&:&~~ \text{D} +\gamma\to {n} + {p},\\
   R6&:&~~ \text{D} +\chi\to {n} + {p} + \chi.
\end{eqnarray}
The signal is the injection of $n$, and neutron detection was the primary objective of the SNO experiment, as a way of resolving the $\nu_e$ deficit from the Sun \cite{SNO:2003bmh}. Therefore we can exploit Canada's investment in heavy water, that is used in both the CANDU nuclear reactors (Canada deuterium-uranium reactors),  and in the SNO experiment. Thus, we are set to evaluate the CANDU-to-SNO scheme for detection, with the signal scaling as the product $R2\times R6 \propto \sigma_{\chi n}^2$. The rate for $R6$ can be calculated in exactly the same way as the textbook example of $R5$ \cite{LL4_QED_1982}.

 %so that SNO would receive an appreciable flux $\chi$ particles which it will then detect via the products of deuteron disintegration. 
 We also consider the flux of $\chi$ particles scattering elastically with spin-dependent nuclei in the ``direct detection" reaction
 \begin{equation}
 R7: ~~ Z + \chi \to Z + \chi,
 \end{equation}
 depositing some energy in a nucleus ``$Z$", which in practice means $^{29}\text{Si}$ or $^{73}\text{Ge}$ nuclei. Currently, such detectors are installed near the natural ``light" water reactors. We analyze hypothetical Si- and Ge-based detectors placed around the Bruce power plant that runs a number of CANDU reactors, and detector performance can be assumed to be the same as in Refs. \cite{Ackermann:2025obx,CONNIE:2024pwt}. 
 %In the case of the Silicon and Germanium based detectors we use statistics from the CONNIE and CONUS detectors to constrain the $\chi$ particle's parameters.
 As before, the signal would scale as $R2\times R7 \propto \sigma_{\chi n}^2$, and our goal is to explore whether this scheme offers additional sensitivity.
 A priori it is not clear whether a very large and very clean ``far" detector SNO would offer any advantage over ``near" detectors that are much smaller, and have much higher background rates. 
 
 In what follows, we shall adopt the simplest models that describe SD interactions, which connect its strength to $\sigma_{\chi n}$, and characterize the resulting sensitivity in the two parameter model $\{m_{\chi}, \sigma_{\chi n}\}$. Section II evaluates the rate for reactions $R2$ and $R4$, and 
 calculates the expected fluxes of $\chi$-particles. 
 Section III addresses $R4$, {\em i.e.} deuteron disintegration signal and deduces the overall sensitivity of the CANDU-to-SNO and Solar-to-SNO schemes of $\chi$ detection. Section IV evaluates possible sensitivity through elastic SD scattering in germanium and silicon near detectors. We reach our conclusions in section V. 
 %We only consider spin-dependent $\chi$ particles as a spin-dependent interaction leads to a much larger deuteron disintegration cross section than the spin-independent case. We consider the $\chi$ particles as having a contact-type interaction with nucleons, forgoing any mediator which may further modify constraints. Using this methodology we produce novel bounds on $\chi$ particles of mass of order 1 MeV with $\chi$-proton cross sections below $10^{-31} cm^{2}$ for the case of detection at SNO. For the cases of detection at Silicon and Germanium based detectors we get bounds which do not improve on preexisting bounds for dark matter in this mass range, further motivating considering alternate avenues for DM detection.  
 
 \section{DM pair-production}
 
 Throughout this paper, we are going to stay on the grounds of effective field theory, and parameterize DM-nucleon interaction via contact operators. 
 The most straightforward form of such interactions (that conserves $CP$) is given by either the interaction of axial-vector or tensor currents:
\begin{eqnarray}
    \mathcal{L}_{int} &=& -A_{N}^a\times (\bar{\chi}\gamma^{\mu}\gamma^{5}\chi)(\bar{N}\gamma^{\mu}\gamma^{5}N),\\
      \mathcal{L}_{int} &=& -A_{N}^t\times \frac12 (\bar{\chi}\sigma^{\mu \nu}\chi)(\bar{N}\sigma^{\mu\nu}N) .
\end{eqnarray}
The DM field $\chi$ can be a Majorana or Dirac fermion, and throughout this paper we adhere to the latter choice. In the non-relativistic limit, both interactions result in the following 
effective Hamiltonian: 
\begin{equation}
H_{\chi N} =  4A_{N}\times ({\bf S}_\chi \cdot  {\bf S}_N )\times \delta^{(3)}({\bf r}_\chi - {\bf r}_N).
\end{equation}
Here ${\bf S}_{\chi,N}$ are the non-relativistic spin operators, ${\bf S} = \frac12 \boldsymbol{\sigma}$. The coupling constants to nucleons can be isospin-dependent, $A_{N} = A_{p},A_{n}$, and have dimension of $-2$, {\em i.e.} similar to a Fermi constant $G_F$. Moreover, it turns out that for the problem considered here, the distinction between axial-vector and tensor nature of the interaction is not important, and we drop a superscript for $A_N$. From this interaction it follows trivially that the non-relativistic elastic scattering cross section of $\chi$ with a nucleon is
\begin{equation}
    \sigma_{\chi N} = \frac{3A_{N}^{2}\mu_{\chi N}^{2}}{\pi}.
\end{equation}
Here $\mu_{\chi N}$ is the reduced mass of the DM-nucleon system, and in this paper we will consider the light DM mass range such that $\mu_{\chi N}\simeq m_\chi$. 
As per common practice for the field, we will use $  \sigma_{\chi N}$ as an input characterizing the strength of the interaction. 

Our goal is to find suitable nucleon-spin-dependent transitions, and recast them as pair-production channels for $\chi\bar\chi$. 
Any nuclear process which results in the emission of a $\gamma$ with energy $E_{\gamma}$ may also produce a $\chi\bar{\chi}$ pair with $2m_{\chi} < E_{\gamma}$. 
In general, we are looking at some initial nuclear state $i$  transitioning to the final state $f$, with emission of an $M1$ $\gamma$ photon 
that can be recast as the SD pair-emission of DM, 
\begin{equation}
i\to f + \gamma(M1)~\to ~ i\to f + \chi\bar\chi({\rm SD}).
\end{equation}
It has to be noted that most of the $\gamma$ emissions in nuclear reactors occur via electric dipole transitions. In particular, one of the most important sources of $\gamma$-emission, ${\rm ^{238}U}(n,\gamma){\rm ^{239}U}$, generates a gamma cascade that is mostly $E1$. Rare $M1$ channels were analyzed in the literature that explores hypothetical axion emission. Ref. \cite{TEXONO:2006spf}, in particular, finds that only $p(n,\gamma){\rm D}$ present a strong source of $\gamma$ with $Q$-value above 2\,MeV. As we are going to see, having a large $Q$-value is very important for both the far and near detection that we consider in this work. Therefore, we concentrate on $R1$ and $R2$ processes that correspond to a relatively large energy release, which of course limits our considerations to heavy water reactors. 

For the capture of thermal neutrons, the energy recoil of the resulting nucleus is very small, and thus the energy released in this process can be treated as entirely going into the $\gamma$ or $\chi\bar{\chi}$ pair. Additionally in this case, both the neutron and nuclei have $v\ll c$ and thus may be treated non-relativistically, allowing us to adopt a semi-relativistic approach to $\chi$ pair production where we treat the nucleons and nuclei non-relativistically but treat the $\chi$ particles, which may have kinetic energy comparable to their mass, relativistically. This amounts to calculating the matrix element as
\begin{equation}
    M = (\bar{\chi}\boldsymbol{\gamma}\gamma^{5}\chi)\bra{\psi_{i}}A_{N}\boldsymbol{\sigma}_{N}\ket{\psi_{f}}
\end{equation}
where $\psi_{i}$ and $\psi_{f}$ are the initial and final nuclear states.

We calculate the rate of $\chi$ pair production in the core region of a ${\rm D_{2}O}$ moderated nuclear reactor. The typical fission rate (number of fission cycles per second) at a nuclear reactor core is\\
\begin{equation}
    \frac{dN_{f.c.}}{dt}(P_{th})= 0.31\times 10^{20}\,{\rm s}^{-1}\times \frac{P_{th}}{1\,\rm GW}.
\end{equation}
where $P_{th}$ is the thermal power of the reactor, normalized here on GW \cite{Xin_2005}. Approximately 2.6 neutrons are released per one fission cycle,
\begin{equation}
    \frac{dN_{n}}{dt}(P_{th}) = 0.8\times 10^{20}\,{\rm s}^{-1}\times \frac{P_{th}}{1\,\rm GW},
\end{equation}
 and are subsequently captured by various nuclei, with some of $n$'s being captured by deuterons in ${\rm D_{2}O}$ moderated reactors. Since the total fraction of neutrons that undergo $R1$ capture is approximately known,  our goal is to find an energy-differential branching ratio of the $R2$ and $R1$ rates,
\begin{equation}
    \frac{d\rm Br_{\chi\bar\chi}}{dE_{\chi}} =\frac{1}{\sigma_{R1}}\times\frac{d\sigma_{R2}}{dE_{\chi}},
\end{equation}
and then use the known occurrence of $R1$ per fission cycle to calculate the $\chi\bar{\chi}$ pair production. 

We calculate the cross sections using the semi-relativistic approach mentioned previously. For the case of the $\gamma$-emitting reaction, the cross section for the capture of thermal neutrons, $\sigma_{R1}$, is dominated by the $M1$ transition and is proportional to $(\mu_{n}-\mu_{p})^{2}$ \cite{Schiff:1937zz}. (This is similar to $n(p,\gamma)\rm D$ reaction, but with the much smaller cross section for $R1$ due to absence of a resonant channel, and extra suppression due to near-orthogonality of the space parts of $i$ and $f$ wave functions \cite{Schiff:1937zz}.) Since the $\chi$-nucleon interaction mirrors the $M1$ transition interaction, this implies that $\sigma_{R2}$ follows a similar form, $\sigma_{R2}\propto (A_{n}-A_{p})^{2}$. An additional important consideration is that the de Broglie wavelength of the emerging DM particles are much larger than the typical nuclear scale of a few femtoseconds that determines the nuclear matrix elements in the $R2$ reaction rates, and therefore, effectively, the small $p_{\chi(\bar\chi)}$ momentum expansion of nuclear matrix elements can be used, and we pick the leading, $0^{\rm th}$ order, term.  Applying this method we get:
\begin{equation}
    \frac{d\rm Br_{\chi\bar\chi}}{dE_{\chi}} = \frac{3(A_{n}-A_{p})^{2}(E_{\chi}E_{\bar{\chi}}-m_{\chi}^{2})p_\chi p_{\bar\chi} % \sqrt{E_{\chi}^{2}-m_{\chi}^{2}}\sqrt{E_{\bar{\chi}}^{2}-m_{\chi}^{2}}
    }{2\pi^{3}(\alpha/m_p^2)(\mu_{n}-\mu_{p})^{2}Q^{3}}.
\end{equation}
Here $p_{\chi} = \sqrt{E_{\chi}^{2}-m_\chi^2}$ and similarly for $\bar\chi$;
$E_{\bar{\chi}} = Q+T_{n}-E_{\chi}$ and $T_{n}$ is the kinetic energy of thermal neutrons, which is much smaller than other energy scales involved and can be safely neglected. Finally, the denominator contains the combination of neutron and proton magnetic moments that determine the $R1$ rate, 
with $\mu_{p}=2.79,\, \mu_{n}=-1.91$ being dimensionless numbers in our notation. Note that the spectrum is fully symmetric with respect to $\bar \chi-\chi$ interchange.

We would like to parameterize this result in terms of the non-relativistic scattering cross section of $\chi$ scattering on a proton. 
Introducing a reference cross section $\sigma_0$ for convenience, 
\begin{equation}
\sigma_0 = \frac{\alpha(\mu_p-\mu_n)^2}{m_p^2}=7.1\times 10^{-29}\,{\rm cm}^2,
\end{equation}
we rewrite the result for the differential branching ratio as 
\begin{equation}
    \frac{d\rm Br_{\chi\bar\chi}}{dE_{\chi}} = \left(\frac{A_{n}}{A_{p}}-1\right)^{2}\frac{\sigma_{\chi p}}{\sigma_0}
  \times \frac{1}{2\pi^2} \times \frac{Q^2}{m_\chi^2} \times \frac{dF(E_\chi)}{dE_\chi},
\end{equation}
where we abbreviated the distribution over energy as 
\begin{equation}
\frac{dF(E_\chi)}{dE_\chi} = \frac{1}{Q^5} (E_\chi E_{\bar\chi}-m_\chi^2) p_\chi p_{\bar\chi} .
\end{equation}

\begin{figure}[h]
\includegraphics[width=7cm]{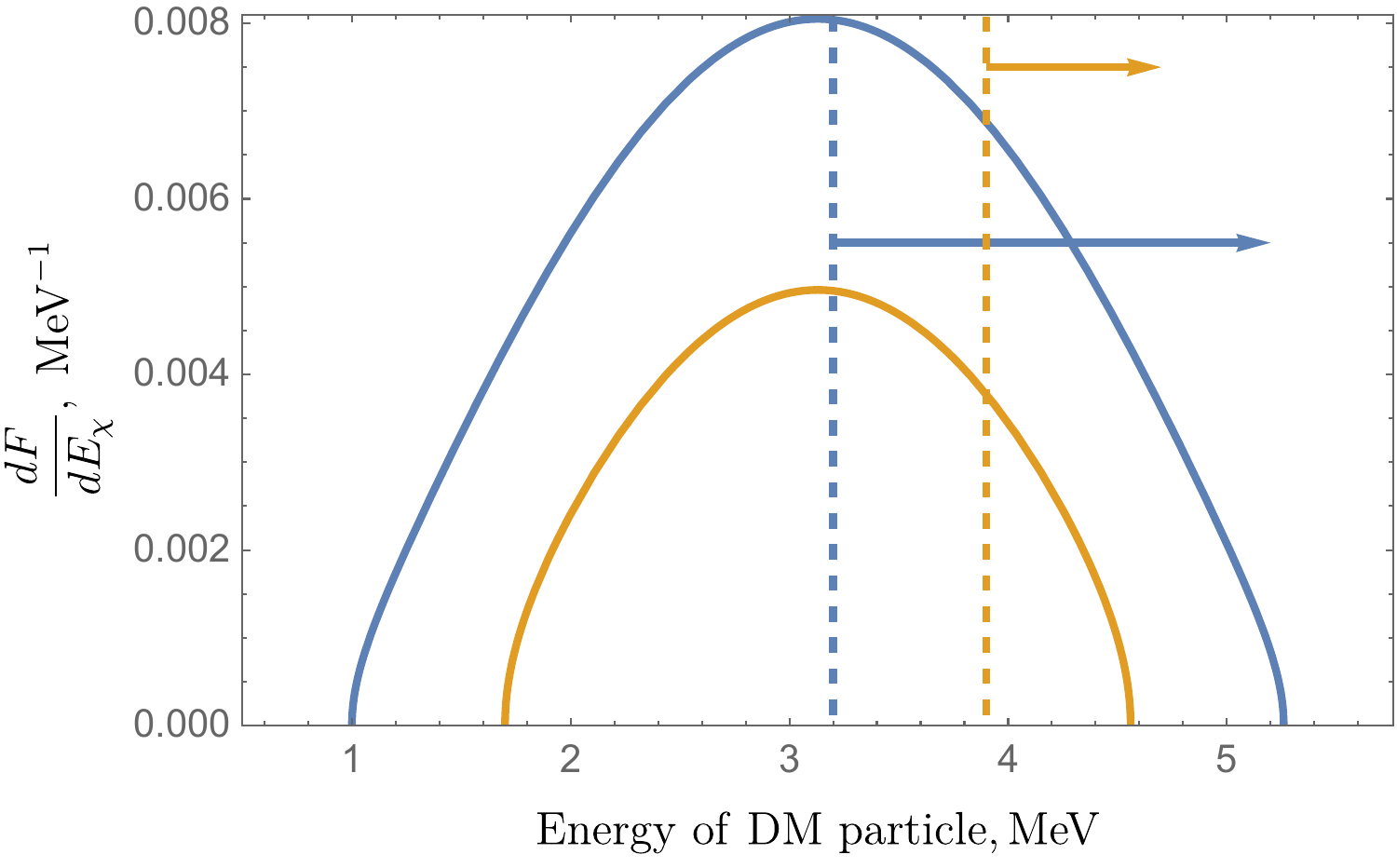}
% Here is how to import EPS art
\caption{\label{fig:2} Energy distribution $\frac{dF}{dE_{\chi}}$ for two representative values of DM mass, $m_{\chi} = 1\,\rm MeV$ (blue) and $m_\chi = 1.7\,\rm MeV$ (brown). The vertical dashed line of the corresponding color indicates the placement of the energy threshold for a subsequent deuteron disintegration, from $E_\chi> E_{\rm thr}=m_\chi +2.2\,\rm MeV$ until the endpoint at $E_\chi = Q-m_\chi$. }
\end{figure}

For the rest of this paper we will take $\frac{A_{n}}{A_{p}} = \frac{\mu_{n}}{\mu_{p}}$, and the numerical result 
for the differential branching of $R2$ can be written as 
\begin{equation}
\frac{d\rm Br_{\chi\bar\chi}}{dE_{\chi}} = 8\times 10^{-4} \times \frac{\sigma_{\chi p}}{10^{-32}\rm\,cm^2}
  \times \left( \frac{1\,\rm MeV}{m_\chi} \right)^2  \frac{dF(E_\chi)}{dE_\chi},
\end{equation}
 Typical energy distributions for several values of $m_\chi$ are shown in Fig.\,\ref{fig:2}.
The total branching ({\em i.e.} the probability of $R2$ for each occurrence of $R1$) is related to the integrals under these curves. For a representative value of 1\,MeV mass and $\sigma_{\chi p}=10^{-32}\rm\,cm^2$, this corresponds to $\rm Br_{\chi\bar\chi} \simeq 1.8\times 10^{-5}$. 

While the cross section of neutron capture on deuteron is, of course, well known,  the rate of neutron capture on deuterons in a realistic reactor setting is not a widely reported quantity. To this end, we estimate the capture rate using the amount of tritium production per year in a CANDU type nuclear reactor. A typical design CANDU reactor produces around 130\,g of $^3{\rm H}$ per year and produces a thermal power of around 2.38 GW\,\cite{PEARSON20181140}. Using these values, the  typical fission rate is $2.25\times 10^{27} \,{\rm yr}^{-1}$ and the amount of tritium produced per year is $2.60 \times 10^{25}\,{\rm yr}^{-1}$, resulting in a rate of $\kappa = 0.0115$ neutron captures on $\rm D$ per fission cycle. Using this rate, our differential flux of $\chi$ and $\bar\chi$ particles is calculated to be 
\begin{equation}
    \frac{d\Phi_{\chi+\bar\chi}}{dE_{\chi(\bar \chi)}}  =2\times \kappa \times \sum\limits_{i} \frac{dN_{n}}{dt}(P_{i}) \times \frac{1}{4\pi L_{i}^{2}}\times\frac{d \rm Br_{\chi\bar\chi}}{dE_{\chi}}
\end{equation}
where we sum over the flux of all CANDU reactors close enough to the point of detection to have an appreciable flux. The factor of two in front originates from the equal fluxes of $\chi$ and $\bar\chi$. 

For example, choosing $L_i$ to be a typical distance to the SNO detector, and $\sum P_i$ to be a typical combined thermal power output, we arrive at the following estimate of the flux
\begin{eqnarray}
    \Phi_{\chi+\bar\chi}(m_\chi=1\,\rm MeV)\simeq 8\times 10^{-2} \,\rm cm^{-2}s^{-1}\\\nonumber \times  \frac{\sigma_{\chi p}}{10^{-32}\rm\,cm^2} 
    \times \frac{\sum P}{20\,\rm GW}\times \left(\frac{250\,\rm km}{L}\right)^2.
    \label{Phi_r}
    \end{eqnarray}
    While this is many orders of magnitude smaller than {\em e.g.} galactic DM flux, it is many orders of magnitude more energetic, and indeed capable of producing an observable signal in the neutrino or DM detectors. 
    
\begin{figure}[h]
\includegraphics[width=7cm]{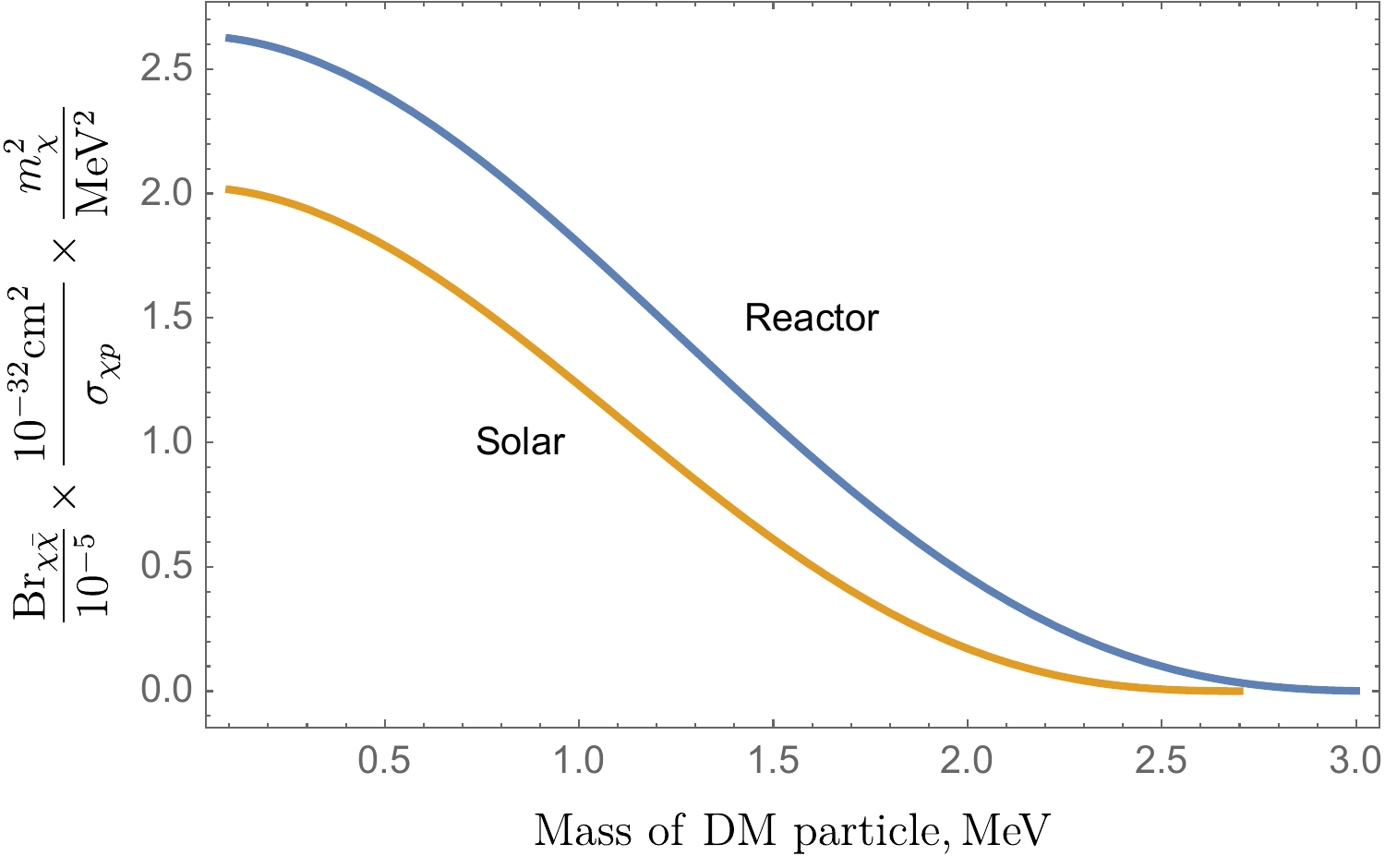}
% Here is how to import EPS art
\caption{\label{fig:Br_rs} Mass dependence of branching ratios to $\chi\bar\chi$ pairs. Due to the normalization on the non-relativistic $\sigma_{\chi p}$ cross section, ${\rm Br}_{\chi\bar\chi}\to \infty$ as $m_\chi\to 0$, and this artificial enhancement has been ``detrended" on this plot via an $m_\chi^2/{\rm MeV}^2$ multiplier. }
\end{figure}

    We can also evaluate the solar $\chi$ flux from reaction $R4$. As is well known, the second step in the $pp$ energy generation chain is the deuteron conversion into $^3{\rm He}$, reaction $R3$. Depending on the energy, the reaction may occur due to a $M1$ or $E1$ transition. We  take, in accordance with the review \cite{Adelberger:2010qa} and references therein, that the $M1$ transition rate in the Sun dominates over $E1$. Conservatively, we take that 80\% of the solar $R3$ reaction is due to $M1$, {\em i.e.} $\kappa^\odot\simeq 0.8$.  The strategy then for determining the $R4$ rate is exactly the same as before, with the only difference being a slightly smaller $Q$-value. This way we arrive at the total solar $\chi$ and $\bar\chi$  flux
    \begin{eqnarray}
    \Phi_{\chi+\bar\chi}^\odot\simeq 2\kappa^\odot\times \Phi^{pp}_{\sum \nu}\times \rm Br^\odot_{\chi\bar\chi},
    \end{eqnarray}
    where $\Phi^{pp}_{\sum \nu}\simeq 6\times 10^{10}{\,\rm cm^{-2}s^{-1}}$ is the predicted (and measured) flavor-inclusive flux of neutrinos from the initial $p+p\to \nu+e^+ + {\rm D}$ process. The solar rate for this process and for $R3$ are exactly equal. Adopting the same assumptions and the same normalization point as before, and calculating  $\rm Br^\odot_{\chi\bar\chi}\simeq 1.2\times 10^{-5} $ numerically, we arrive at the estimate of the flux
     \begin{eqnarray}
     \label{Phi_s}
    \Phi_{\chi+\bar\chi}^\odot({m_\chi \!=\!1\,\rm MeV}) \simeq 1.1\times 10^{6}{\,\rm cm^{-2}s^{-1}}  \frac{\sigma_{\chi p}}{10^{-32}\rm\,cm^2} .
    \end{eqnarray}
    While this flux appears to be much larger than the reactor flux (\ref{Phi_r}), one should keep in mind that in order to avoid the downgrade in energy due to solar opacity, one would need to have a cross section several orders of magnitude smaller than reference point for $\sigma_{\chi p}$ used in (\ref{Phi_s}). In conclusion, in this section we provide the $m_\chi$ dependence of the branching to $\chi\bar\chi$ pairs, {\em i.e.} the ratio of rates $R2/R1$ and $R4/R3$ for the ${\rm D_2O}$ reactors and the solar $pp$ cycle. These $m_\chi$-dependent branching ratios are plotted in Fig.\,\ref{fig:Br_rs}.
    
\section{Detection at SNO}
The SNO experiment resolved the solar neutrino problem by detecting the rate of the neutral current (NC) interaction of $^8$B neutrinos on deuterons, which is insensitive to the specific neutrino flavor. The process ${\rm D}(\nu,\nu)np$ is visible because of the $\gamma$ associated with the subsequent neutron capture. If on top of the confirmed neutrino signal there exists a boosted-DM-induced deuteron break-up, ${\rm D}(\chi,\chi)np$, its rate can be limited from the overall consistency of the NC rate with the predictions of the solar model and the self-consistency of three active neutrino flavor oscillation model. Thus, the total probability per time of the deuteron disintegration by NC and $\chi(\bar\chi)$, 
\begin{equation}
\frac{1}{N_{\rm D}}\frac{d N^{\rm tot}_{\rm D\!\!\!\!/}}{dt} = \int dE_\nu \frac{d \Phi_{\nu\odot}}{dE_\nu}\sigma_{\nu\rm D\!\!\!\!/}+\int dE_\chi \frac{d \Phi_{\chi+\bar\chi}}{dE_\chi}\sigma_{\chi\rm D\!\!\!\!/}\,,
\end{equation}
one can require the second term to be no more than the $10\%$ of the first one \cite{SNO:2011hxd} (see also Refs. \cite{Bahcall:2004pz,Butler:2000zp,Nakamura:2002jg} for relevant calculations of flux and cross section), setting the maximum allowed rate for ${\rm D}(\chi,\chi)np$ as 
\begin{equation}
\label{limiting_rate}
\frac{1}{N_{\rm D}}\frac{d N^{\chi}_{\rm D\!\!\!\!/}}{dt} = \int dE_\chi \frac{d \Phi_{\chi+\bar\chi}}{dE_\chi}\sigma_{\chi\rm D\!\!\!\!/}< 3\times 10^{-37}\,\rm s^{-1}.
\end{equation}

Next we calculate the cross section of deuteron dissociation by $\chi$-particles using the semi-relativistic approach analogous to the calculations from the previous section. The physics of the deuteron break up is well known and occurs via the relative nucleon spin flip, aided by the low-lying resonance in the $S=0$ channel, ${\rm D\to D^*}\to np$. (Deuteron disintegration due to exotic neutrino properties was discussed earlier in {\em e.g.} Refs.\,\cite{Grifols:2004yn,Pospelov:2011ha}.) Following the textbook approach \cite{LL4_QED_1982}, we derive 
\begin{eqnarray}
d\sigma_{\chi \rm D \!\!\!\!/} =\frac{1}{4\pi^2} (A_n-A_p)^2  m_{\chi} ^2 \times \frac{\kappa dp dq}{k^2q}\\
\times J^2 \times \left(1 +\frac{E_iE_f}{m_\chi^2} 
\mp\frac{E_f^2+E_i^2-q^2-2m_\chi^2}{6m_\chi^2}\right). \nonumber
\end{eqnarray}
Here $E_{i,f}$ and $k_{i,f}$ are the energies and spacial momenta of the $\chi$ particle before and after the scattering, $q$ is the spatial momentum transfer, $q = |\vec{k}_f -\vec{k}_i|$, $p$ is the relative momentum of $n$ and $p$ in the final state, and $\kappa$ is the ``zero-radius" parameter of the deuteron ground state wave function, $\kappa = \sqrt{m_p I}$, where $I=2.2$\,MeV is the deuteron binding energy. 
$J$ parametrizes the spatial overlap $\bra{\psi_{i}}\ket{\psi_{f}}$ of the initial and final nuclear states defined as
\begin{eqnarray}
J(q,p) = q \int  \exp(-\kappa r)\sin(pr+\delta) \nonumber\\
\times \exp(i \frac12 q r \cos\theta_r)d\cos\theta_rdr.
\end{eqnarray} 
Phase $\delta$ relates to the energy of the virtual $\rm D^*$ level, $\cot\delta = \sqrt{m_pI_1}/p$ with $I_1\simeq 0.067$\,MeV. The range of the integration is defined as $k_i-k_f\leq q\leq k_i+k_f$ and $ 0\leq p^2/m_p\leq E_i-I-m_\chi$. Finally, the $\mp$ sign is due to axial or tensor generalization of the spin-spin interaction. In practice, the influence of this term is at most at a few percent level, and can be neglected. 

\begin{figure}[h]
\includegraphics[width=7cm]{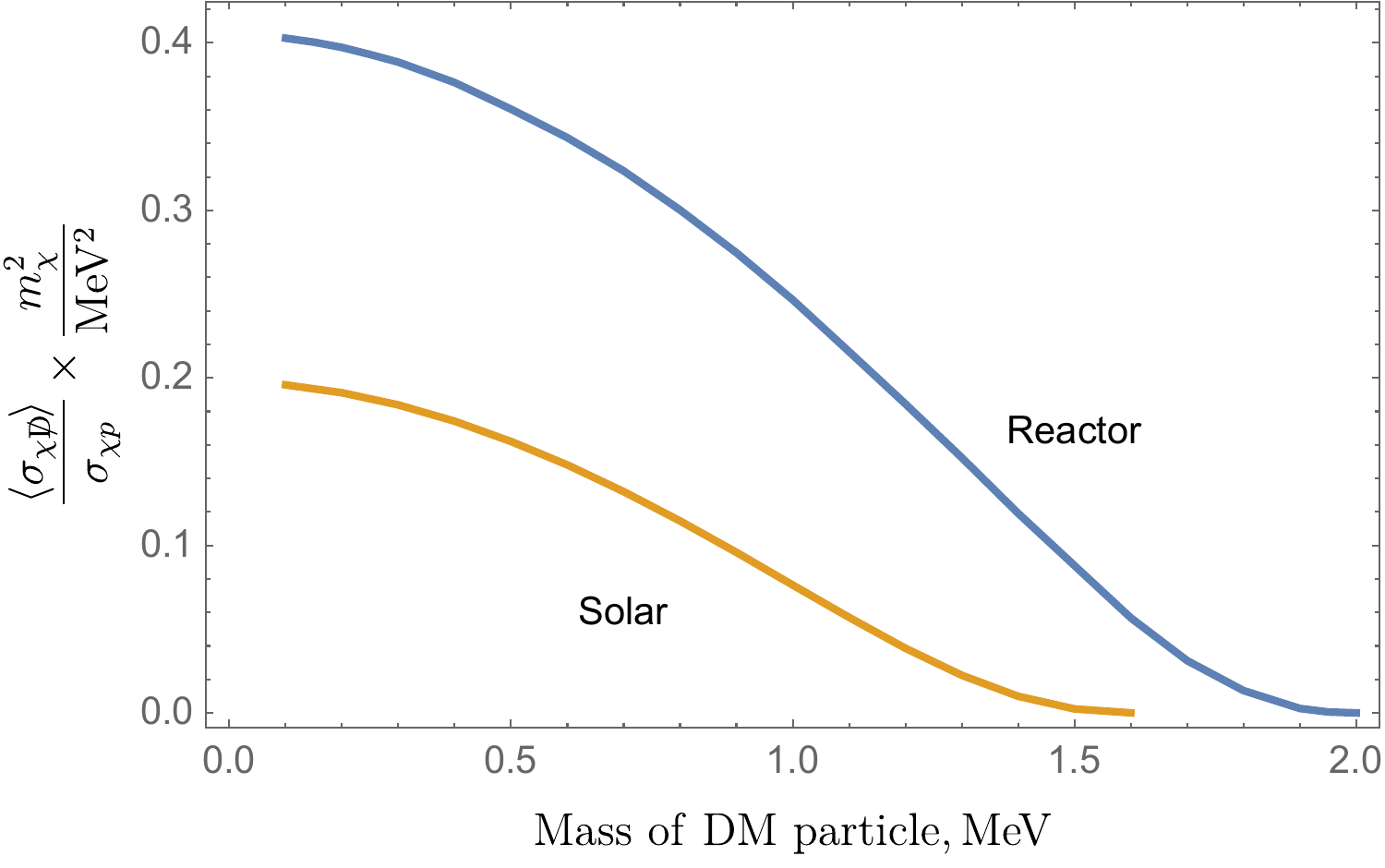}
% Here is how to import EPS art
\caption{\label{fig:NEW} Flux-averaged deuteron disintegration cross sections $\langle \sigma_{\chi\rm D\!\!\!\!/} \rangle$ normalized on $\sigma_{\chi p} ({\rm MeV}/m_\chi)^2$. These curves are similar in shape to those of Fig.\,\ref{fig:Br_rs}, with a smaller $m_\chi$ at endpoints due to the energy threshold of $R6$.}
\end{figure}

The expression for $J(q,p)$ can be easily found in an analytic form, and the rest of the integral is performed numerically. To this end, we would like to make the same choice as the previous section, $A_n/A_p = \mu_n/\mu_p$, and trade $A_p^2$ for $\sigma_{\chi p}$. It is convenient to express the results as the {\em flux-averaged} cross section, 
\begin{equation}
\langle \sigma_{\chi\rm D\!\!\!\!/} \rangle \equiv \frac{\int_{E_{\rm thr}}  \sigma_{\chi\rm D\!\!\!\!/}(E_\chi) \frac{dF(E_\chi)}{dE_\chi}dE_\chi  }{{\int_{\rm full}}   \frac{dF(E_\chi)}{dE_\chi}dE_\chi } .
\end{equation}
Fig. \ref{fig:NEW} shows the $\langle \sigma_{\chi\rm D\!\!\!\!/} \rangle/\sigma_{\chi p}$ ratio for two flux shapes, reactor and solar. 
For $m_\chi=1$\,MeV, the deuteron disintegration cross section relative to free scattering $\sigma_{\chi p}$ reaches 25\% for the reactor and 7.6\% for the solar flux spectra ($A_p/A_n=\mu_p/\mu_n$ is assumed). 

To calculate the $\chi+\bar\chi$ flux at SNO's location, we use the information about CANDU reactors at Bruce, Pickering, and Darlington Power stations. Importantly, all reactors were in place by the time the main scientific runs of the SNO experiment. We will abbreviate them as ``B", ``P", and ``D", and use the following input for distances and thermal power \cite{IAEA_PRIS_Canada_2024}: 
\begin{eqnarray}
P_{\rm B} = 22.7\,{\rm GW},~~L_{\rm B} =2.4\times 10^7\,{\rm cm};\nonumber\\
P_{\rm P} = 14.0\,{\rm GW},~~L_{\rm B} = 3.4\times 10^7\,{\rm cm};\\
P_{\rm D} = 11.7\,{\rm GW},~~L_{\rm D} =3.5\times 10^7\,{\rm cm}.\nonumber
\end{eqnarray}
With that input, and assuming that reactors stay operational 90\% of time, we get a more precise flux value compared to (\ref{Phi_r}),  
\begin{eqnarray}
    \Phi_{\chi+\bar\chi}\simeq 0.14 \,{\rm cm^{-2}s^{-1}}\times  \frac{\text{Br}_{\chi\bar\chi}(m_\chi)}{1.8\times 10^{-5 } }.
    \label{Phi_r}
    \end{eqnarray}
    
    Multiplying this flux by the calculated branching ratio, Fig.\,\ref{fig:Br_rs}, and by the flux-averaged cross section, Fig.\,\ref{fig:NEW}, we derive the $\chi$-induced rate for the deuteron disintegration. Comparing this to the maximum allowed rate (\ref{limiting_rate}), one concludes that the SD cross section of $\chi$-nucleon scattering for 1\,MeV DM is limited better than 
    \begin{equation}
    \sigma_{\chi p}(m_\chi = 1\,\rm MeV) <  2.8\times 10^{-34}\,{\rm cm^{2}},
    \end{equation}
    while the full mass range of the constraint is shown in Fig.\,\ref{fig:money_plot}. The slanted shape of the exclusion line is the direct consequence of the simplified model chosen, as well as the non-relativistic normalization of the cross section, and was noted in previous works such as Ref.\,\cite{Dent:2019krz}. 
    Repeating the same calculations for the solar flux, we can derive the sensitivity to even smaller values of $\sigma_{\chi p}$,
  \begin{equation}
  \label{rel_nrel}
    \sigma_{\chi p}(m_\chi = 1\,\rm MeV) <  1.8\times 10^{-37}\,{\rm cm^{2}}.
    \end{equation}
    
While this bound appears to be impressively strong, one should keep in mind the {\em ceiling} for the solar bounds associated with solar opacity. 
For this paper we estimate the cutoff cross section, based on the following simplified picture of $\chi$ propagation. A single event of an elastic collision with protons will lead to a small loss of kinetic energy in the $\chi$ particle emerging from the ${\rm D}$-$p$ fusion reactions in the solar core, while the cross section of $\chi$-$p$ scattering is larger than non-relativistic one:
\begin{eqnarray}
\label{DeE}
\Delta E_\chi = - \frac{\langle q^2 \rangle}{2m_p} \sim - O(1)\frac{E_\chi^2 - m_\chi^2}{m_p},\\
\sigma_{\chi p }(E_\chi) = \sigma_{\chi p} \times \frac{E_\chi^2+m_\chi^2}{2m_\chi^2}
\end{eqnarray}
The coefficient in (\ref{DeE}) can be calculated, and it has a slight dependence on the axial vector or tensor choice for spin-spin interactions. (In the limit of $E_\chi \gg m_\chi$, this coefficient is $\frac{10}{9}\simeq 1.1$ for the axial vector choice of interactions.) 
A single collision will not lead to more than a sub-\% energy loss, but will result in the total re-orientation of the momentum. In the limit of a large number of collisions, this will lead to the diffusion of particles out of the solar core, so that the total distance traveled would scale as $N\ell_{\rm col}$, and distance from the center of the Sun as $\sqrt{N} \ell_{\rm col}$, where $\ell_{\rm col}$ is the collision length, and $N$ is the number of collisions. Simple estimates show that at $m_\chi \le 1$\,MeV, the {\em maximum } number of collisions that $\chi$ particles can tolerate without falling in energy below $I_b+m_\chi$ threshold is close to 80-100 collisions. 

In order to find the position of the maximum cross section that does not completely cuts off the flux of energetic particles, we simulate, via a random walk method, the propagation of $\chi$ particles with emission points given by the predicted distribution of $pp$ events. Using the collision length at the center of the Sun as an input, we determine that for $\ell_{{\rm col},r=0} = 5\times 10^{-3}R_\odot$, 1\% of emitted particles will have less than 80 collisions, and therefore carrying enough energy to dissociate deuterons. The number of collisions is extremely sensitive to the input and at $\ell_{{\rm col},r=0} = 0.01R_\odot$ already 15\% of particles have fewer than 80 collisions, and 1\% of particles less than 20. For $\ell_{{\rm col},r=0} = 3\times 10^{-3}R_\odot$, the probability for fewer than 80 collision drops below $10^{-3}$. Therefore, $\ell_{{\rm col},r=0} = 5\times 10^{-3}R_\odot$ reasonably determines the minimum collision length and the maximum cross section that can be probed, which we conservatively evaluate at the highest energy point, $Q-m_\chi$, where $Q=Q(R4)=5.49$\,MeV. Taking the number density of protons in the core to be $n_p \simeq 3\times 10^{25}\,{\rm cm}^{-3}$, we translate $\ell^{\rm min}_{{\rm col},r=0} = 5\times 10^{-3}R_\odot$ condition into 
\begin{eqnarray}
\sigma^{\rm max}_{\chi p}(E_\chi = Q-m_\chi) \simeq 10^{-34}\,{\rm cm}^2 \\
\to ~\sigma^{\rm max}_{\chi p}(m_\chi=1\,\rm MeV) \simeq 10^{-35}\,{\rm cm}^2.\nonumber
\end{eqnarray}
Here the second line specifies the maximum non-relativistic cross section for the reference point of $m_\chi=1$\,MeV. We can observe that solar emission coupled with reaction $R6$ in SNO excludes a range of relevant cross sections $\sigma_{\chi p}$, Fig.\,\ref{fig:money_plot}. Cross section $\sigma^{\rm max}_{\chi p}$ determines the upper boundary of the red region. 

Note that recently Ref.\, \cite{Ge:2024cto} proposed to constrain spin-independent $\chi$-nucleon scattering using the $\chi\bar\chi$ production in the Sun followed by recoil inside some future advanced direct detection devices. Current direct detection experiments do not produce any viable limits. Indeed, as we found, the opacity of the Sun requires cross sections to be less than $10^{-35}\,{\rm cm}^2$, which is currently unattainable for the direct detection detectors in this recoil range. We note in passing that the SI interaction in the $10^{-36}\,{\rm cm}^2$ can be constrained by the SNO experiment via the $E1$-type emission and deuteron break-up. 
    
% to constrain $,m_{\chi}$ and $\sigma_{\chi p}$ which is given in Fig.~\ref{fig:3}.

\section{Direct detection at near sites}
The reactor flux of stable exotic particles as seen by the SNO detector at $L_{\rm far}\sim250$\,km distance can be enhanced, by many orders of magnitude, at near sites. Many experiments performed in the vicinity of large commercial reactors are placed at a typical distance of $L_{\rm near}\sim25$\,m, resulting in the following enhancement:
\begin{equation}
\label{flux_enhancement} 
\frac{\Phi_{\chi, \rm near}}{\Phi_{\chi, \rm far}}\propto \frac{L_{\rm far}^2}{L_{\rm near}^2}\times \frac{P_n}{\sum P}\propto 10^7,
\end{equation}
where we took into account that a flux at the far location is ``cumulative", while the flux at a near location is by definition set by a single reactor with power $P_n$.
At the same time, much smaller size of near detectors, and much higher background levels do not allow taking a full advantage of this enhancement in terms of the reach to fundamental parameters, such as $\sigma_{\chi p}$ in this paper. Translating, for example, the limiting count rate (\ref{limiting_rate}) to slightly different units, we arrive at the event rate $1.5\times 10^{-6}$/day/kg of heavy water. Counting rates at the CE$\nu$NS detector are at best (in the reactor ``on" and ``off" difference), $O(1)$/kg/day in the relevant recoil energy range. Thus, the flux enhancement is largely offset by far worse detection sensitivity. It is also clear that SD scattering, considered in this paper, does not have the coherent enhancement of the CE$\nu$NS process, and therefore only a small, sub-10\,\% fraction of nucleons in a Si- or Ge-based detector will participate in the $\chi$-nucleus scattering. In this section we evaluate sensitivities of possible direct detection/CE$\nu$NS detectors placed in close proximity to a CANDU reactor. 

In both germanium and silicon based detectors, the SD detection mechanism ($R7$) comes from the  recoil of the nuclei after scattering with the $\chi$ particle. Since our $\chi$ particles have significantly less energy and mass than the mass of the nuclei, we may calculate the elastic scattering cross section of $\chi$ particles with these nuclei taking these nuclei as being in the large mass limit. In the case of both the $^{29}\text{Si}$ and $^{73}\text{Ge}$ nucleus the $\chi$ particle largely couples with the valence neutron. The matrix element that relates nuclear spin with the matrix element of $\boldsymbol{\sigma}_n$ contains some additional suppression that can be determined with the help of nuclear theory, {\em e.g.}  via modeling of the magnetic moments for the two nuclei. In the case of $^{29}{\rm Si}$ this suppression factor is $\kappa_N\simeq 0.25$ and in the case of $^{73}{\rm Ge}$ this factor is 0.8 ~\cite{Ressell:1993qm}. Following this approximation the elastic scattering cross section between $\chi$ and either of these nuclei is
\begin{equation}
   \sigma_{\chi N} = \frac{3\kappa_{N}^{2}A_{n}^{2}(E_{\chi}^{2}+m_{\chi}^{2})}{2\pi},
\end{equation}
where $\kappa_{N}$ is the corresponding ``neutron spin content" of nuclear spin. The corresponding parameterization in terms of $\sigma_{\chi {p}}$ is then
\begin{equation}
 \sigma_{\chi N} =  \sigma_{\chi {p}}\times \frac{\kappa_{N}^{2}({\mu_n}/{\mu_p})^2(E_{\chi}^{2}+m_{\chi}^{2})}{2 m_{\chi}^{2}},
\end{equation}
where again proportionality of $A_{n,p}$ to $\mu_{n,p}$ is assumed.

For the analysis of sensitivity to SD scattering at a near site, we will consider a hypothetical situation of Si and Ge detectors placed at a distance of 30 meters from the core of one of the Bruce power reactors. We shall assume a similar quality of data to that of the CONNIE and CONUS detectors for the silicon and germanium based detectors respectively. (We assume that a Si detector is based on Skipper-CCD technology that has achieved very low detection thresholds.) We will also assume natural isotopic composition, with known concentrations of $^{29}{\rm Si}$ and $^{73}{\rm Ge}$. Isotopic enrichment could, in principle, boost their relative concentrations by up to a factor of ten.

For both nuclei, the limit $m_{\chi}\ll m_{N}$ is certainly valid, and the maximum recoil energy of the nucleus is 
\begin{equation}
    T^{max}_N \simeq  \frac{2(E_{\chi}^{2}-m_{\chi}^{2})}{m_{N}}.
\end{equation} 
Notice that this energy is always limited by $2Q^2/m_N$, where $Q=6.26$\,MeV is the energy release of $R2$. In that sense, the recoil energy is {\em less} than for CE$\nu$NS signal where $E_\nu$ endpoint is at $\sim 8$\,MeV. Evaluating the expected counting rates for a $1\,$MeV mass particles, we get for germanium detectors:
\begin{eqnarray}
\label{Ge_counts}
{\cal R} = 4.93\times 10^{2}\,\frac{\rm events}{\rm kg\times day} ~~~~~~\nonumber\\ \times \left(\frac{\sigma_{\chi p}}{\rm 10^{-32}\,cm^{-2}}\right)^2\times
\left(\frac{ P_n}{2.5\,\rm GW}\right)\times \left(\frac{25\,\rm m}{L}\right)^2,\\
T^{max}_{\rm Ge}= 0.8\,\rm keV,~~~~~~~~~~~~~ \nonumber
\end{eqnarray}
and for silicon detectors, 
\begin{eqnarray}
{\cal R} = 1.5\times 10^{2}\,\frac{\rm events}{\rm kg\times day} ~~~~~~\nonumber\\ \times \left(\frac{\sigma_{\chi p}}{\rm 10^{-32}\,cm^{-2}}\right)^2\times
\left(\frac{P_n}{2.5\,\rm GW}\right)\times \left(\frac{25\,\rm m}{L}\right)^2,\\
T^{max}_{\rm Si} = 2\,\rm keV.~~~~~~~~~~~~~  \nonumber
\end{eqnarray}
The smallness of recoil energy creates the biggest challenge for this scheme: even though (\ref{Ge_counts}) is much larger than the limiting rates of \cite{Ackermann:2025obx}, dominant number of predicted events will fall under the 
Ge detection threshold. This energy is further reduced by the so-called quenching factor, when converted to equivalent electron ionization. For germanium detector, this quenching factor is on the order of $0.15$ in the relevant sub-keV energy range. 

\begin{figure}[!b]
\includegraphics[width=7cm]{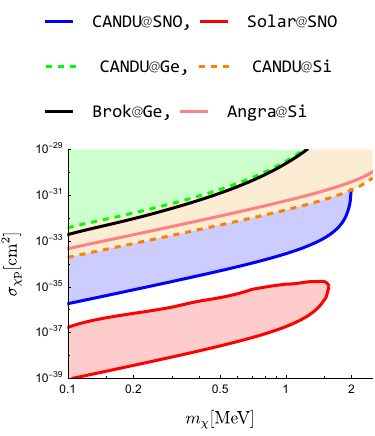}% Here is how to import EPS art
\caption{Excluded regions on  $\{m_{\chi},\sigma_{\chi p}\}$ parameter space, which result from the CANDU produced $\chi$ counting rate at SNO (SNO) and the solar produced $\chi$ counting rate at SNO (Solar). {\em Possible} sensitivity reach from Silicon (Si) and Germanium (Ge) nuclear recoil experiments placed near a CANDU reactor are also shown by dashed lines. Existing constraints from Ge and Si detectors at natural water reactors are shown by solid lines. }
\label{fig:money_plot}
\end{figure}

The majority of events, especially for Ge detector, will be below the detection threshold, but their number just below the threshold can be very large. 
Therefore, in order to have a more realistic estimate of the signal, 
for both such detectors we need to consider the energy resolution of the signal. For both types of detectors the energy resolution follows a gaussian with standard deviations of 48 eV and 34 eV for the CONUS and CONNIE detectors respectively ~\cite{Ackermann:2025obx,CONNIE:2019xid,CONNIE:2024off}. Including this energy resolution factor, we calculate the predicted event rates.
% we need to include an additional factor of the probability of measuring a signal above threshold to calculate the counting rate per nucleus, resulting in the new expression
%\begin{equation}
 %   r =\int dE_{\chi}\frac{d\Phi}{dE_{\chi}}\sigma_{\chi\text{N}}P_{N}(E_{\chi}).
%\end{equation}
We then compare this theoretical rate to the reactor-on minus reactor-off rate for the corresponding signal energy range above threshold for the corresponding detector to constrain $m_{\chi}$ and $\sigma_{\chi\text{p}}$. The resulting constraints, or more precisely sensitivity limits, for the silicon detector and the germanium detector are given in Fig.\,\ref{fig:money_plot}. In both cases the resulting constraints are poor because the $\chi$ particles lack enough mass and energy to provide the nuclei with enough recoil energy that would be detectable. 

In order to use the {\em existing} experiments at natural water reactors, more in-depth treatment of DM pair production via ${\rm ^{56} Fe}(n,\chi\bar\chi){\rm ^{57} Fe}$ is needed, along with other energetic $M1$-type transitions. We estimate the pair-production of $\chi\bar\chi$ pairs in the reaction of neutron capture on iron following the general prescriptions for axion emission outlined in Ref. \cite{Avignone:1988bv}, and applied recently to ${\rm ^{56} Fe}(n,a){\rm ^{57} Fe}$ reaction in Ref. \cite{Waites:2022tov}. The occurrence of ${\rm ^{56} Fe}(n,\gamma){\rm ^{57} Fe}$ reaction is estimated using the analysis in Ref. \cite{TEXONO:2005fmk} performed for a commercial high-power natural pressurized water reactors. The resulting constraints are included in Fig.\,\ref{fig:money_plot}.

\section{Discussion and Conclusions}

In this paper, we have evaluated four types of constraints on light spin-dependent dark matter scattering on nucleons. In the global effort to discern the nature of dark matter, one of the main directions is the lowering of detection thresholds, which allows for lighter and lighter DM particles to be probed. There are many challenges on the way, starting from precipitously rising backgrounds, as thresholds become lower. SD scattering poses an additional challenge as the coherent enhancement by the number of nucleons is absent. The main goal of our paper was to set independent constraints using already available data from the neutrino experiments, such as SNO. The main mechanisms of DM production considered in this paper are nuclear reactors, and solar production in the $pp$ cycle of energy generation. 

In this paper, we derive novel constraints on the SD interaction, starting at the effective field theory level involving nucleons. The generalization of non-relativistic spin-spin interaction to the product of axial-vector or tensor currents is possible, and as we were able to see that it does not lead to widely different predictions. This leads us to believe that bounds derived here are indeed widely applicable. If the models are specified even further, {\em e.g.} to the level of quarks, one could employ more data such as meson decays. There, the difference between tensor and axial-vector interactions will lead to drastically different outcomes. For our treatment, it suffices that parameters $A_{p,n}$ do not have strong form factors, and are approximately equal between $s$- and $t$- ({\em i.e.} production and scattering) channels for $\chi \bar \chi$. Moreover, for an MeV-scale DM, we consider processes with momentum transfers limited to $O(5)$\,MeV and even for a massless mediator one does not expect strong $q$ dependence of the form factors or drastic differences between scattering and pair-production. This is in contrast with other approaches that, for example, try to compare DM scattering and DM pair-production at TeV colliders where extrapolations from $t$- to $s$-channel is problematic. Furthermore, it is well known that if the spin-independent scattering is resolved via an almost massless mediator $A_\mu$ coupled to vector currents (such as $\bar\chi \gamma_\mu \chi$, $\bar e \gamma_\mu e$ etc), the $t$-channel scattering cross section can be many orders of magnitude larger than the $s$-channel annihilation and/or pair production. In the SD case the situation is different, as {\em e.g.} $ A_\mu $ coupling to the axial vector current will lead to the strong enhancement of $\chi\bar\chi$ pair-emission/annihilation (see {\em e.g.} Ref. \cite{Dror:2017nsg}), so that the dominance of scattering cross section over the $s$-channel will no longer be applicable. This lands further support to the statement that the derived limits will apply even if $A_{p,n}$ are resolved as vector or axion-type scalar exchange. 

In conclusion, we would like to summarize main findings of our paper: 
\begin{itemize}
\item The fortuitous placement of SNO detector in reasonable proximity to heavy water CANDU reactors in Ontario, Canada, allows us to consider ${\rm D}+n$ fusion as a source of $\chi\bar\chi$ pairs ($R2$) that are subsequently detected via deuteron break-up at SNO ($R6$). The relatively strong energy release in the neutron capture by deuterons allow us to place limits on DM particles with masses up to $2$\,MeV. These limits are in the $10^{-34}\,{\rm cm}^2$ domain, and indeed are much stronger than the most advanced direct detection experiments with lowest thresholds can offer.  

\item The solar emission due to the isospin mirror reaction $R4$ can be evaluated in a similar manner resulting in the flux of $\chi$ particles from the Sun that can be limited with the use of the same SNO results. Importantly, the solar $\chi$ flux has a ceiling, or the cross section beyond which the solar particles produced in the core will lose too much energy at the exit from the Sun. This sets an independent {\em non-overlapping} constraint on solar $\chi$ emission. The range of excluded cross sections contains a $\sigma_{\chi p} \sim 10^{-36}\,{\rm cm}^2$, which is of special interest in WIMP physics.

\item Even though heavy water reactors do not have widely known CE$\nu$NS detectors nearby, we have considered a hypothetical case of low-threshold Ge- and Si-containing detectors near a CANDU reactor zone as possible way of constraining SD interactions with direct detection technology. Our findings show that this scheme does not offer a promise to surpass CANDU-to-SNO constraints. The sensitivity could be enhanced if one makes a special effort to create a detector dominated by the spin-containing nuclei, but even then the DM signal is expected to be subdominant to CE$\nu$NS. 

\end{itemize}

A gap left between solar and reactor SNO limits, Fig.\,\ref{fig:money_plot}, begs the question of how one can probe it. Perhaps one way of achieving it is to have a ``miniaturized" version of the SNO detector (perhaps with the mass of several hundred kg) placed in a reasonable proximity to a nuclear reactor. If indeed counting rates as low as $10^{-2}$/kg/day can be achieved (from $\chi$-induced spallation of $n$) while benefiting from flux enhancement similar to (\ref{flux_enhancement}), the parameter space in the gap can be probed/closed. 

To conclude, we would like to point out that the spin-independent interactions, not considered in our paper, can also be limited in a similar way. In that case there are many more production channels, and the near detectors looking for elastic recoil would benefit from nucleon coherence factors. The analysis of such possibilities is postponed for future work.

%We have shown that a spin-dependent DM-nucleon leads to a significant flux of $\chi$ particles produced a $D_{2}O$ moderated nuclear reactors. Subsequent scattering of this DM in the SNO detectors sets novel constraints on $\sigma_{\chi p}$ parameterized as $\{m_{\chi}, \sigma_{\chi p}\}$ which fall below previous constraints in this region. In comparison, subsequent scattering of this DM in silicon and germanium based detectors fail to produce novel constraints due to the DM lacking enough mass and enough to produce a substantial enough signal. Our results expand upon previous constraints on the parameter space of spin-dependent DM. Further refinements of our constraints can be achieved within this model by using a fully relativistic treatment to obtain the relevant scattering cross section.

{\bf Acknowledgments.} The authors are
supported in part by U.S. Department of Energy Grant
No. DE-SC0011842. MP would like to acknowledge useful conversations with Dr. H. An on related topics.

\clearpage
\bibliography{apssamp}% Produces the bibliography via BibTeX.

\end{document}